\documentclass[9pt,twocolumn]{extarticle}
\pdfoutput=1  
  
\usepackage{soul}
\usepackage{titlesec} 
\usepackage{comment}
\usepackage[backend=bibtex,style=nature]{biblatex}
\DeclareNameAlias{sortname}{first-last} 
\DeclareNameAlias{default}{first-last}
\bibliography{biblio} 
\usepackage{dsfont} 
\usepackage{graphicx}
\usepackage{subcaption}
\usepackage[margin=0.9in]{geometry}   
\usepackage[usenames,dvipsnames]{color}
\usepackage[colorlinks,linkcolor=Blue,urlcolor=Blue,citecolor=Blue]{hyperref}
\usepackage{amsmath,amssymb,braket}
\usepackage{amsfonts}
\usepackage{bbm}
\usepackage[squaren]{SIunits}
\usepackage{pifont}

\usepackage{array}
\newcolumntype{P}[1]{>{\arraybackslash}p{#1}}
\newcolumntype{Q}[1]{>{\centering\arraybackslash}p{#1}}

\usepackage{diagbox}

\usepackage{mathpazo}

\usepackage{courier}
\normalfont

\usepackage[T1]{fontenc}
\usepackage{caption}  
\usepackage{upgreek}

\newcommand{\ad}[1]{\textsuperscript{#1}\kern-2pt}

\usepackage{amsmath,amsthm,mathtools}

\makeatletter
\def\blx@maxline{77}
\makeatother

\usepackage{capt-of}

 \usepackage{flushend}

\usepackage[ruled]{algorithm2e} 
\usepackage{newfloat,algcompatible} 
\def\mytitle{
A programmable topological photonic chip 
\vspace{-2mm}}      
 
\title{\vspace{-1cm}\Huge\textbf{\textrm{\mytitle}}}  
 
\author{Tianxiang Dai$^{1,8\star}$, Anqi Ma$^{1,8}$, Jun Mao$^{1,8}$, Yutian Ao$^{2,3}$, Xinyu Jia$^{1}$, Yun Zheng$^{1}$, Chonghao Zhai$^{1}$, Yan Yang$^{4\star}$, \\Zhihua Li$^{4}$, Bo Tang$^{4}$,  Jun Luo $^{4}$, 
Baile Zhang$^{2,3}$, Xiaoyong Hu$^{1,5,6,7\star}$, Qihuang Gong$^{1,5,6,7}$, Jianwei Wang$^{1,5,6,7\star}$}
 
\date{} 
\begin{document}
\twocolumn[{
\maketitle 
\vspace{-7mm} 
\begin{center}
\begin{minipage}{1\textwidth}
\begin{center}
\textit{\textrm{
\textsuperscript{1} State Key Laboratory for Mesoscopic Physics, School of Physics, Peking University, Beijing 100871, China 
\\\textsuperscript{2} Division of Physics and Applied Physics, School of Physical and Mathematical Sciences, Nanyang Technological University, Singapore 637371, Singapore
\\\textsuperscript{3} Centre for Disruptive Photonic Technologies, The Photonics Institute, Nanyang Technological University, Singapore 637371, Singapore
\\\textsuperscript{4} Institute of Microelectronics, Chinese Academy of Sciences, Beijing 100029, China
\\\textsuperscript{5} Frontiers Science Center for Nano-optoelectronics \& Collaborative Innovation Center of Quantum Matter, Peking University, Beijing 100871, China 
\\\textsuperscript{6} Collaborative Innovation Center of Extreme Optics, Shanxi University, Taiyuan 030006, Shanxi, China
\\\textsuperscript{7} Hefei National Laboratory, Hefei 230088, China
\\\textsuperscript{8} These authors contributed equally to this work.
\\\ ~~~Emails to: tianxiang.dai@pku.edu.cn, yyang10@ime.ac.cn, xiaoyonghu@pku.edu.cn, jww@pku.edu.cn}}
\end{center}
\end{minipage}
\end{center}

\setlength\parindent{12pt}
\begin{quotation}
\noindent 
{\bf{Controlling topological phases of light  has allowed experimental  observations of abundant topological phenomena and   development of  robust photonic devices.
 The prospect of more sophisticated controls with topological photonic devices for practical implementations  requires  high-level  programmability. 
Here, we demonstrate a fully programmable topological photonic chip with large-scale integration of silicon photonic nanocircuits and microresonators. 
Photonic artificial atoms and their interactions in our compound system can be individually addressed  and controlled, 
therefore allowing arbitrary altering  of structural  parameters and geometrical configurations for the observations of dynamic topological phase transitions and  diverse photonic topological insulators.
By individually  programming artificial atoms on the generic chip, it has allowed comprehensive   statistic 
characterisations of topological robustness against relatively  weak disorders, as well as counterintuitive topological Anderson phase transitions induced by strong disorders. 
Our generic topological photonic chip that can be rapidly reprogrammed to implement multifunctionalities, prototypes a flexible and versatile platform for possible applications across fundamental science and topological technologies. 
}} 
\end{quotation}}]  
 
\noindent

Topological insulators (TIs) have garnered significant interest, 
because of the abundant physical mechanisms underlying non-trivial bands and  potential applications of topological boundary modes~\supercite{Topologicalinsulators,qi2011topological}. 
Since the first discovery of the quantum Hall effect~\supercite{PhysRevLett.45.494, thouless1982quantized}, the intricate diagrams of topological phases have developed as a sprawling tree with intertwined branches, encompassing dimensionality~\supercite{ryu2010topological}, symmetry~\supercite{slager2013space}, non-Hermiticity~\supercite{
feng2017non,Ozdemir2019}, and defects~\supercite{lustig2022photonic,liu2021bulk,Gao2016}. 
One leap  happened when topology met photonics~\supercite{Haldane,moore2010birth,Lu2014,RevModPhys.91.015006,khanikaev2017two}. 
Photonic systems provide numerous advantages for topological physics and technologies, such as %reliable phase stability,
 noise-free environment, few constraints on lattice geometry, large diversity of optical materials, high controllability  of optical devices, and widely adoptable nonlinear optical effects ~\supercite{moore2010birth,Lu2014,RevModPhys.91.015006,khanikaev2017two,khanikaev2013photonic,jurgensen2021quantized,slobozhanyuk2017three}. 
Topological photonics  that is initially proposed as an extension of topological materials in optical artificial structures, is  emerging as an independent field and is revolutionising  
optical science and technologies. 
%that surpasses the preliminary expection of emulating quantum Hall effect by breaking time-reversal symmetry in gyromagnetic photonic crystals~\supercite{Wang2009}.
For examples, integer quantum Hall TIs~\supercite{Wang2009,Hafezi2011,zilberberg2018photonic}, quantum spin Hall TIs~\supercite{Khanikaev2013,bliokh2015quantum}, Floquet TIs~\supercite{Rechtsman2013,PhysRevX.3.031005},
high-order TIs~\supercite{benalcazar2017quantized,,el2019corner}, non-Hermitian TIs~\supercite{wang2021generating,zhao2019non} and many other interesting  topological phenomena have been  observed  in various photonic systems. 
Practical topological devices, e.g, topological  delay lines \supercite{Hafezi2011},
topological lasers~\supercite{Bandreseaar4005,zeng2020electrically}, %kodigala2017lasing}, 
topological single-photon\supercite{Mittal2018,BlancoRedondo568} and entangled-photon sources~\supercite{Mittal2021,dai2022topologically}, and topological devices for  communications~\supercite{yang2020terahertz,kumar2022phototunable}, have been intensively developed and explored.

Those observations of topological effects and demonstrations of topological devices are  reported on a large variety of optical devices with specifically  designed periodic structures or geometries. 
It is essential  to flexibly and precisely  control  topological phases of light in  programmable topological photonic devices at both levels of fundamental and applied science. 
First,  the dynamics of topological phase transition (TPT) relies on strong reconfiguring of  structural parameters of the devices. Topological invariants  maintain until bands cross 
so that a dramatic altering of parameters is required. In typical measurements, TPTs are observed in several different devices, or even a  joint multivariate effort is necessary for TPTs~\supercite{Zhaxylyk,weidemann2022topological,liu2022topological}. 
 Though TPTs are possible by globally tailoring the devices with an adoption of nonlinear  effects~\supercite{maczewsky2020nonlinearity,xia2021nonlinear,nonlinear,}, or mechanical displacement~\supercite{Cheng2016},  portraying TPTs by more direct and accurate  approaches is demanded. 
Individually  programming each artificial atom as well as the atom-atom interactions may represent the ultimate control of the system.  This however remains  challenging  in many natural and artificial  topological systems, so as in photonics.  
 Second, most of previous observations of topological phenomena rely on static analysis of single or several samples. 
 Comprehensively certificating topological robustness by statistical measurements, and 
probing interesting statistical topological phenomena such as topological Anderson insulators~\supercite{stutzer2018photonic,li2009topological,meier2018observation} and amorphous topological insulators~\supercite{mitchell2018amorphous,jia2023disordered}, require the ability  to individually program artificial atoms and their interactions so as to control disorders. Fabricating a large number of samples  with precisely controlled disorders for such  statistical analysis is   impractical. 
Third, as topology in matters derives from the collective behavior of atoms in the lattice, the  geometry  of lattice determines the interrelationships between neighboring atoms  and  the overall topological properties. 
Topology of bands varies in  dimensions~\supercite{ryu2010topological}, and lattices with various geometries also make different symmetries~\supercite{slager2013space}, resulting in TIs in  different classes. 
Previous investigations on TIs in diverse lattices however  rely on completely different samples, which necessarily needs custom design and fabrication of samples.

\begin{figure*}[ht!] 
  \centering
  \includegraphics[width=0.8\textwidth]{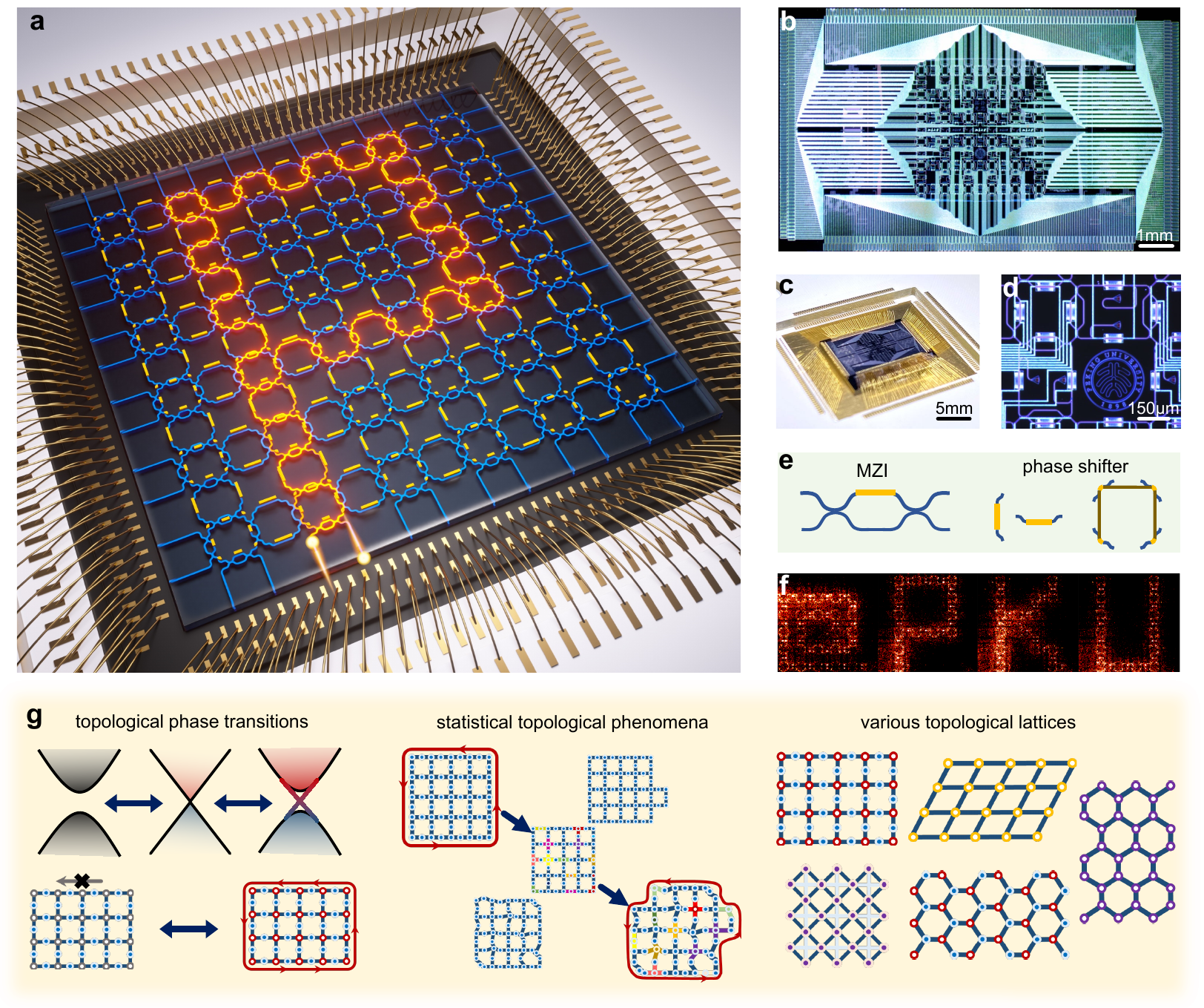} 
  \caption{\textbf{A fully programmable topological photonic chip.}  
  \textbf{a}, Conceptual diagram.  It  integrates large-scale photonic nanowaveguide circuits and microring resonators. In total, 96 microrings with high quality factors are regularly positioned in a 6$\times$6 square lattice. 
  All rings (artificial atom) can be individually controlled by integrated thermo-optical phase shifters (gold parts), achieving arbitrary resonant phases in all rings, phase differences between the two paths of link rings and coupling strength between neighboring rings. 
At the boundaries, 24-in-by-24-out ports are connected to the lattice.
\textbf{b}, Photograph of a fabricated topological chip. The silicon chip is fabricated using CMOS processes and it monolithically integrates 2712 components in a 11mm $\times$ 7mm footprint, including 408 low-loss directional couplers, 300 phase-shifters with 528 thermal isolators, 48 grating couplers for optical access and  120 tapping  ports for light field imaging, 600 electronic access and 708  transmission lines. 
\textbf{c}, Photograph of a packaged chip. The chip is wire bonded  on a multi-layer printed circuit board (PCB). 
External  electronic drivers with a number of 600 channels are used to individually control  300 phase shifters. % are established between chip pads and PCB by . %Fiber arrays for optical coupling are temporally removed to avoid obstructing the view. 
\textbf{d}, Optical microscopy image of a three-ring unit cell.  
\textbf{e}, Diagrams of reconfigurable optical components, including MZIs and phase shifters. 
\textbf{f}, Imaging of real-space distributions of electromagnetic field. As examples, the chip is flexibly reprogrammed to display "@PKU" respectively.
\textbf{g}, The generic chip is reprogrammed to implement multifunctionalities: 
dynamic topological phase transitions,  observation of statistical topological phenomena, and benchmarking of TIs in various lattices.}
  \label{fig:concept} 
\end{figure*} 

\begin{figure*}[ht!] 
  \centering
  \includegraphics[width=0.912\textwidth]{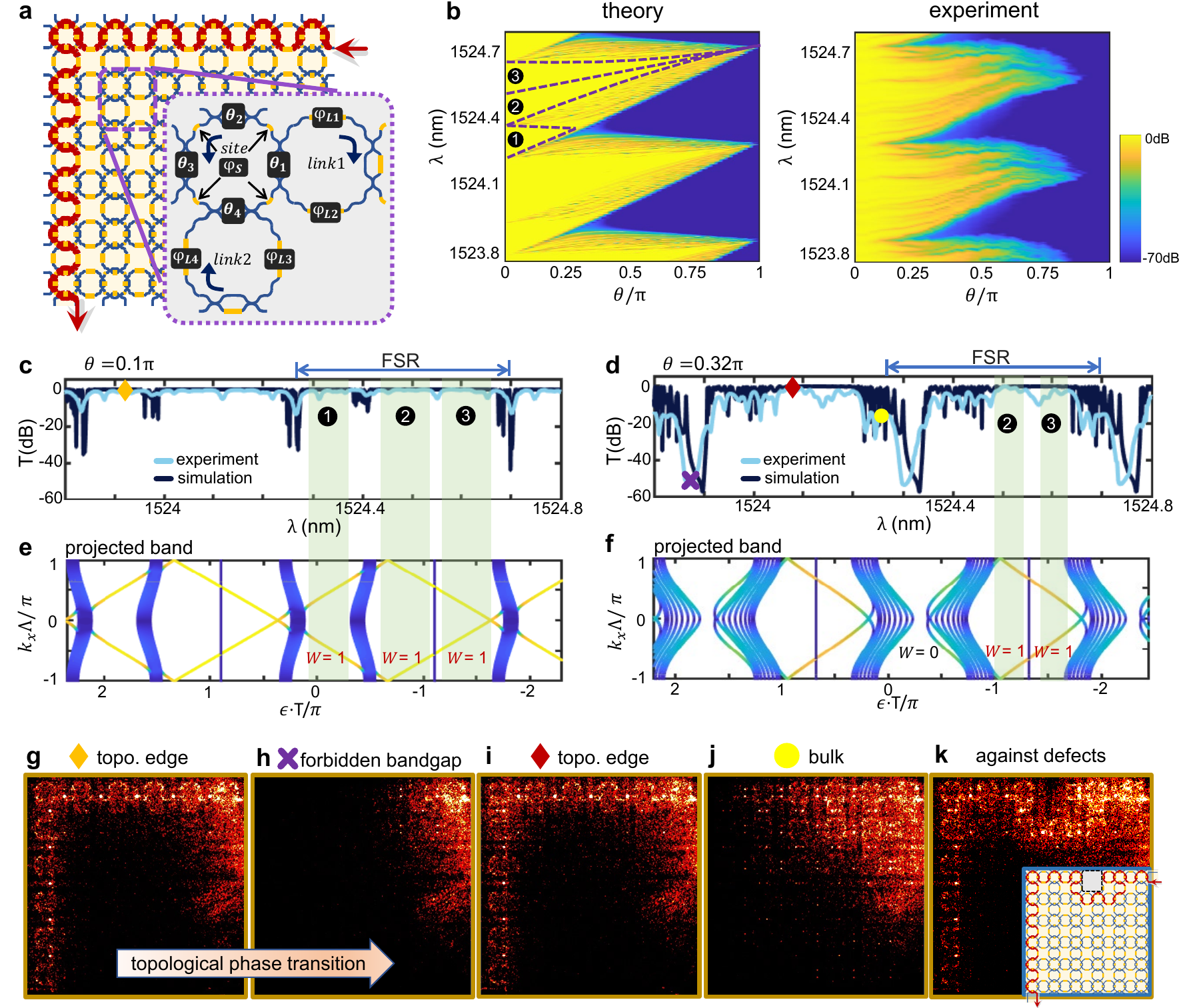} 
  \caption{\textbf{Coupling-strength-controlled TPTs in Floquet TIs.}  
  \textbf{a}, A three-ring model for Floquet TIs. Using 9 parameters in a single unit cell, it fully describes the quasi-energy bandstructure. %The ports for excitation and detection and a schematic "S"-shape path of topological edge modes are illustrated. 
  \textbf{b},  TPTs driven by coupling strength. Transmission spectra as functions of wavelength $\lambda$ and the parameter $\theta$ on coupling strength are shown, where $\theta$ is negatively correlated with the coupling strength and the amplitude transmittance of a MZI is $\cos(\theta/2)$.  
   % Theoretical (left panel) and experimental (right panel)  in wavelength $\lambda$ and $\theta$. 
The boundaries of nontrivial bandgaps in one FSR are indicated by purple dashed lines. Theoretical results (left panel) are in good agreement with experimental results (right panel).
% A typical feature is observed that 
Boundary states at bandgaps \ding{182} disappear with a continuous variation of $\theta$ near the critical point at $\theta=0.272\pi$, while edge modes in bandgaps \ding{183} and \ding{184} exist throughout the entire range of $\theta$ variation. %And the degenerate bulk modes between them are excited in experiment because of tiny imperfections.
  The attenuation of light for large $\theta$ in experiment is due to resonant enhancement in rings, which increases the effective optical length so as  loss. 
  \textbf{c,d}, Measured spectra at $\theta=0.1\pi$ (strong coupling) and $\theta=0.32\pi$ (weak coupling), and \textbf{e,f}, their calculated bandstructures, respectively. 
  The windows of edge modes are visually enhanced. 
  \textbf{g-j}, Imaged real-space distributions of electromagnetic field under different points marked in spectra in ({c,d}): TPT from ({g}) topological edge modes to ({h}) forbidden bandgaps at bandgap \ding{182}, ({i}) edge modes at bandgaps \ding{183} in weak coupling regime, (\textbf{j}) randomly distributed bulk mode from the non-degenerate bulk bands.
  \textbf{k}, A boundary cell in FTI is removed by adjusting its coupling to the "bar" state, which forms a lattice defect. High-transmission topological edge modes bypass the hole and present its robustness against atomic vacancies.
  Note that on the link ring paths, we tapped out -35dB light using diffractive grating couplers for better imaging of light fields, which results in the appearance of regularly distributed bright spots. Noises at the top right come from light reflection from input fiber.}
  \label{fig:thetaTPT} 
\end{figure*}

\begin{figure*}[ht!]  
  \centering
  \includegraphics[width=1.0\textwidth]{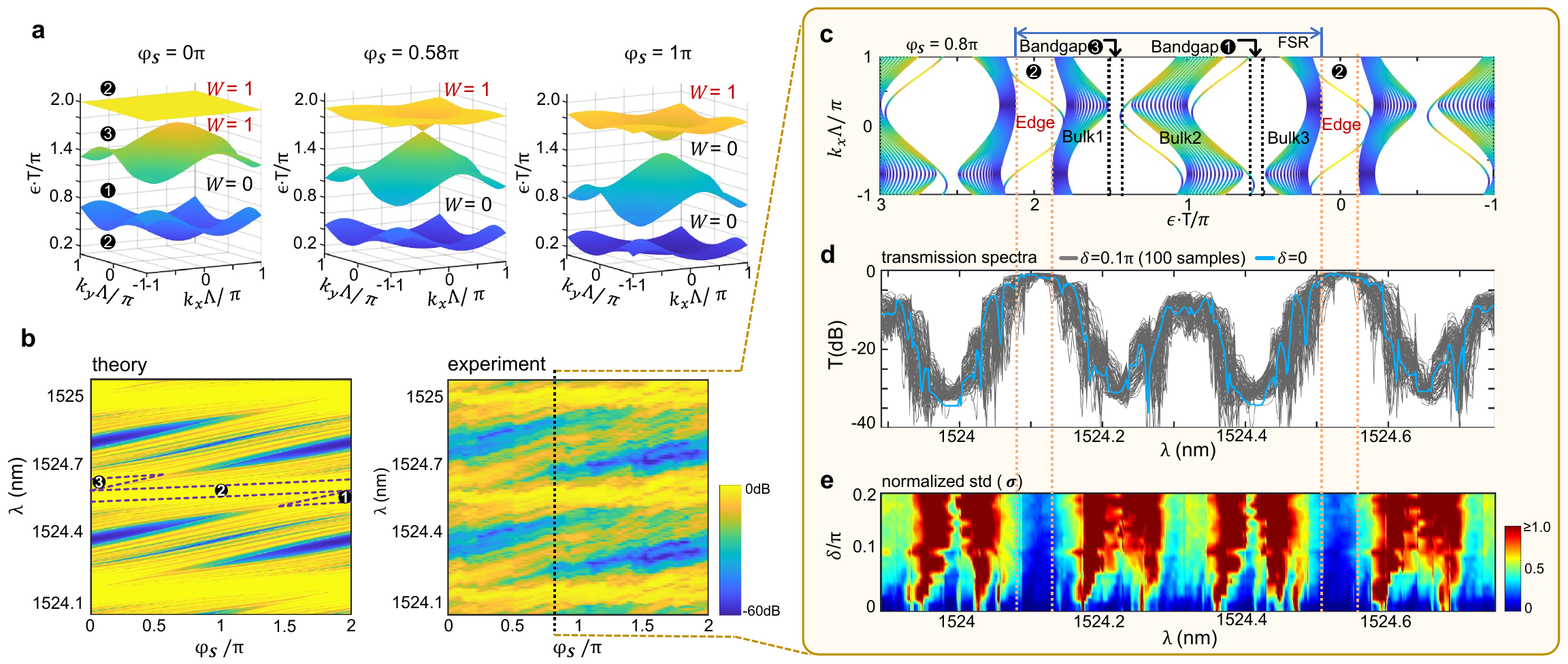} 
  \caption{\textbf{Resonant-phase-controlled TPTs and statistic verification of topological robustness.}  
  \textbf{a}, Calculated bandstructures in $\varphi_{S}$-controlled TPT. 
  Starting from the weak coupling regime ($\theta = 0.4\pi$), by increasing $\varphi_{S}$ from 0 to $\pi$, site rings and link rings become detuned, reaching a maximum detuning at $\varphi_{S}=\pi$. TPT happens at bandgap \ding{184} when $\varphi_{S}=0.58\pi$, making it a trivial forbidden bandgap. 
  \textbf{b}, Theoretical (left panel) and experimental (right panel) transmission spectra as functions of wavelength $\lambda$ and phase $\varphi_{S}$.   As a global phase shift is introduced by $\varphi_{S}$, TPT happens at bandgap \ding{182}. 
The mapping of bandgaps changes from \{\ding{182}, \ding{183}, \ding{184}\} to \{\ding{184}, \ding{182}, \ding{183}\} after a 2$\pi$-evolution of phase $\varphi_{S}$. 
  \textbf{c},  Calculated projected bandstructures at $\varphi_{S}=0.8\pi$. 
  It is plotted as a reference to demonstrate the robustness of topological edge modes. % against phase disorders. 
  \textbf{d} and   \textbf{e},  Experimental  verification of topological robustness with statistic measurements by individually controlling the phase disorders in all rings. 
In (d), a sets of 100 samples with uniformly distributed random phases are chosen in the range of $\delta\times[-0.5,0.5]$ and  $\delta=0.1\pi$ in measurement. Measured transmission spectra for disordered device are shown as the gray background, and the spectrum for an ideal device without disorders is plot  as a blue line.  
In the topological edge modes, the flat plateaus with high transmission are slightly influenced, while in all other regimes, severe broadening and small dips owing to obstruction from random local modes appear. The measured standard deviation (normalised) of transmission spectra under different strength of disorders are plot in (e). Evident windows with low fluctuations are exactly corresponding to the regimes of topological edge modes. Standard deviations are colour coded and the key is provided at the right bottom.
 }
  \label{fig:TPTphi} 
\end{figure*}

\begin{figure*}[ht!] 
  \centering
  \includegraphics[width=0.9\textwidth]{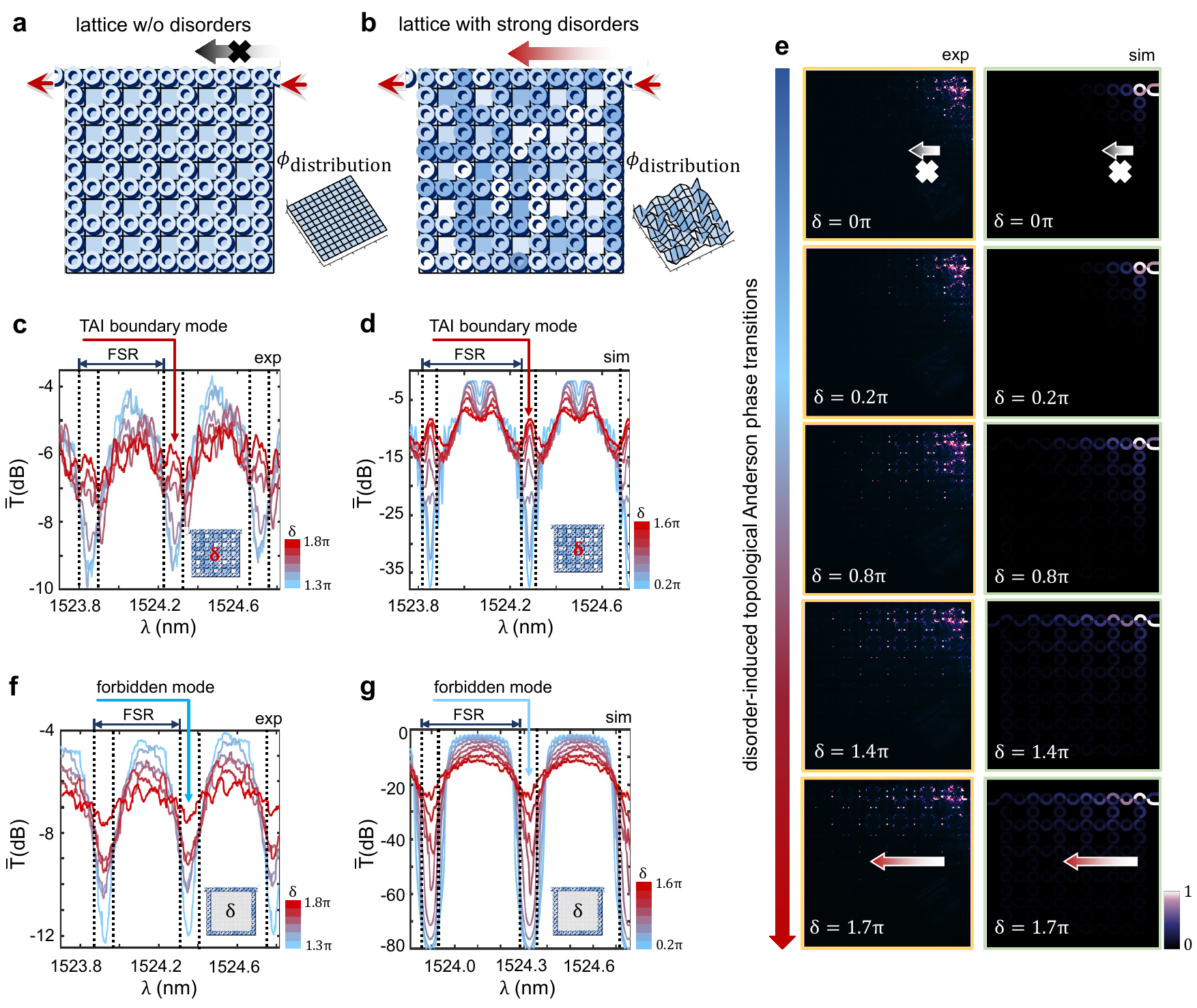} 
  \caption{\textbf{Observation of topological Anderson phase transitions with statistical measurements by individually programming the TI with strong disorders.}  
  \textbf{a}, Ideal Floquet TI in weak coupling regime($\theta=0.3\pi$) with no disorder. All microrings are matched to resonance. There is a forbidden bandgap within one FSR. 
 \textbf{b}, TAI induced by strong disorders.  Topological Anderson phase transitions happen at the forbidden bandgaps and connect them with TAI boundary modes. 
  Statistic measurements of the TAI with a large variety of disorders are necessary to observe the topological Anderson phase transitions, which are realised on a single device by individually controlling phase disorders in all rings in our experiment. 
    \textbf{c} and  \textbf{d}, Measured and simulated transmission spectra of the TAI with different disorders. 
    For each disorder, 100 samples are generated on the chip for statistic measurements, and the mean spectra is plotted.  
  As an increase of disorders, an intriguing peak that represents the TAI boundary mode (indicated by the red arrow) gradually emerges at the low-transmission dip where used to be the forbidden bandgaps in the ideal lattices with no disorders. 
 \textbf{e}, Imaging the dynamic process of topological Anderson phase transitions. Each image is an accumulated field distributions of 100 samples. 
 The phase transition from a forbidden mode to a transported TAI boundary mode along the upper boundary is observed. Simulation results are shown for comparsion, which are in good agreement with  experimental results. 
Different to random diffusions of bulk modes, the TAI boundary modes propagate along the boundary and rapidly decay into the bulk. 
 \textbf{f} and  \textbf{g}, Measured and simulated transmission spectra of a 1D trivial device. The shape of transmission spectra does not change with increasing disorders, and the low-transmission dip  corresponding to the forbidden bandgaps remained a dip (indicated by the blue arrow), i.e, no phase transition occurred. 
}

  \label{fig:TAI} 
\end{figure*}

\begin{figure*}[ht!] 
  \centering
  \includegraphics[width=1 \textwidth]{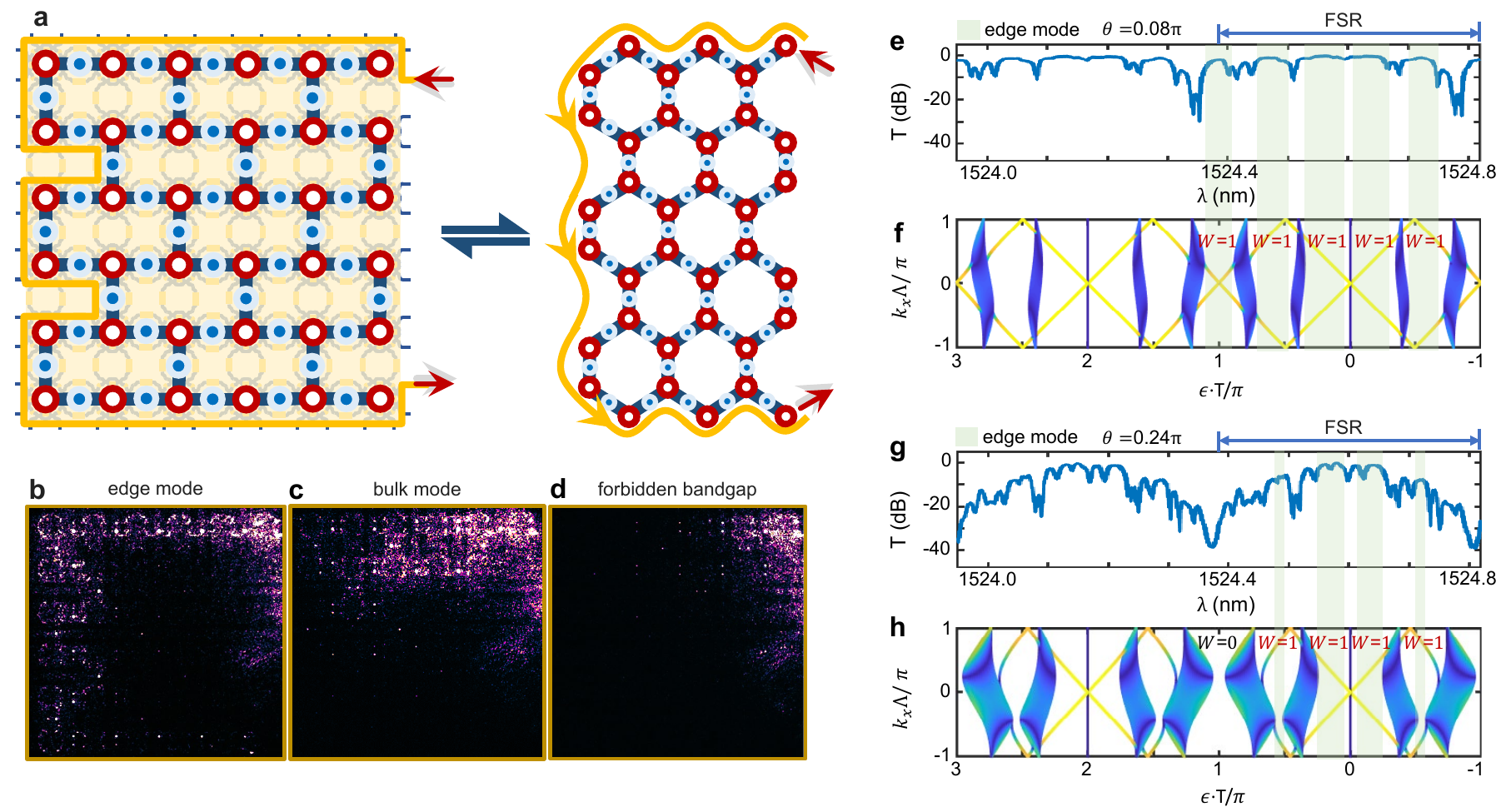} 
  \caption{\textbf{Floquet topological insulators in the honeycomb lattice by reconfiguring the lattice geometry.}  
  \textbf{a}, A reconfigured square lattice that is equivalent to the honeycomb lattice. 
  \textbf{b-d}, Measured real-space distributions of light field in different modes: ({b}) topological edge mode, ({c}) bulk mode, and ({d}) forbidden bandgap. 
  \textbf{e} and \textbf{f}, Measured transmission spectrum and calculated  projected band when the device works in $\theta=0.08\pi$. There are five bulk bands in one FSR, distinct from the three-particle model in a square lattice.
   The winding numbers $\mathcal{W}$ in all bulk bandbaps are one, implying the existence of non-trivial boundary modes. 
   \textbf{g} and \textbf{h}, Measured transmission spectrum and calculated projected band when the device works in $\theta=0.24\pi$.  
   When decreasing the coupling strength down to the critical point of $\theta=0.19\pi$, TPTs happen at the bandgap across $\epsilon=\pi/T$, turning it a trivial forbidden bandgap ($\mathcal{W}$=0), while other bandgaps remain nontrivial. }

  \label{fig:honeycomb} 
\end{figure*}

In this work, we report a highly   programmable topological photonic chip. The  chip has generically integrated  a lattice of large-scale silicon photonic nano-waveguide circuits and microring resonators, and is fabricated  by   the complementary metal-oxide-semiconductor (CMOS) processes. 
%Utilizing the large-scale integration and mature programmable components of silicon photonics, 
When we consider each ring as an artificial atom, our  photonic chip can be regarded  as an artificial lattice that allows arbitrary individual control of atoms as well as  the coupling strength and hopping phase between atoms. 
The generic  chip can be rapidly reprogrammed to implement different functionalities, such as to 
 dynamically transform topological phases of Floquet TIs, observe statistical topological phenomena (statistical analysis of topological robustness and topological Anderson phase transitions), and realise a diverse class of topological insulators with various  lattices (e.g, 1D Su-Schrieffer-Heeger TIs, and 2D Floquet TIs in square and honeycomb lattices). 
Our work prototypes a flexible, versatile and instant-reprogrammable topological photonic platform.

Figure~\ref{fig:concept} illustrates   an overall concept. 
% introduction to the device. 
The photonic topological insulating chip is  devised on a two-dimensional lattice of coupled-microring resonators and nanophotonic circuits, as shown in Fig.\ref{fig:concept}a. Each microring emulates one atom and the photonic  chip emulates the artificial atom lattice. In experiment, we realised a $6 \times 6$ square lattice, embedding in total  96 microrings, each of which has intrinsic quality factors  in the order of $10^{5}$. 
As shown in   Fig.\ref{fig:concept}e, resonance of each microring can be individually controlled, and the coupling between microrings (both strength and phase) can be arbitrarily controlled by Mach-Zehnder interferometers (MZIs) with ultra-high extinction ratio of about 50dB. 
The device operates at the wavelength of 1525 nm. 
One fabricated and packaged chip is shown in Figs.\ref{fig:concept}b-d. 
The high-level controllability and programmability of generic photonic chip enable 
sophisticated implementations of dynamic topological phase transitions, statistical topological processes, and diverse topological lattices, see Fig.\ref{fig:concept}g. 
As an initial    test of the flexible and fast programmability of the generic chip, Fig.\ref{fig:concept}f shows  imaged field distributions of "@PKU" symbols, and a Supplementary Video shows  real-time modulations of "HELLO" letters. 

 We first report arbitrary controls of the bandstructure of Floquet TIs in  the three-particle model. 
The famous Floquet theory provides an effective temporal approach for TIs with no need of truly breaking time-reversal symmetry~\supercite{rudner2020band,PhysRevX.3.031005}.  Demonstrating the full modulation capability requires comprehensive controls of structural parameters, which remains experimentally  exclusive. 
A zoom-in view of a  three-ring unit is shown in Fig.\ref{fig:thetaTPT}a and the real structure  is shown in Fig.\ref{fig:concept}d.  
By reconfiguring  four parameters on the coupling strength (i.e, $\theta_{1-4}$) and five parameters on the phase (i.e, $\varphi_{L1-L4} ~ \text{and} ~ \varphi_{S}$) in a three-ring unit cell, arbitrary topology in the three-band structure can be constructed based on the Floquet band theory. 
We experimentally characterise  
two types of Floquet TPTs which are driven by the coupling strength ($\theta$) and resonant phase ($\varphi_{S}$), respectively. 

$\theta$-driven topological phase transitions: 
%And we first focus on the former. 
Simultaneously tuning  all coupling parameters of $\theta_{1-4}=\theta$ and across the TPT critical point ($\theta=2\arcsin(\sqrt{2}-1)$ $\approx$ $0.272\pi$), the bandgaps close and reopen, resulting in  disappearance of topological edge modes (indicated by \ding{182} in Fig.\ref{fig:thetaTPT}b) and the phase transition at bandgap \ding{182} from an topological phase to trivial phase.
The topological invariant winding number ($\mathcal{W}$) is used to explicitly portray the topology (i.e, $\mathcal{W} =1$ for nontrivial phase, while  $\mathcal{W} =0$ for trivial phase). 
Topological invariants also intuitively reflect in transport properties. %After determining the optical I/O ports in Fig.~\ref{fig:thetaTPT}a, 
Figure \ref{fig:thetaTPT}b shows the theoretical and experimental transmission spectra with a fine tuning of $\theta$ from 0 to $\pi$ (i.e, transmittance of MZIs from 1 to 0). The flat and high-transmission regimes  (outlined by dashed lines) indicate topological edge modes in one free spectral range (FSR). One FSR corresponds to one 2$\pi/T$ period in quasi-energy $\epsilon$, where $T$ is the period of Floquet evolution. 
 %which also perfectly match the experimential results. 
Figures \ref{fig:thetaTPT}c and \ref{fig:thetaTPT}d show two measured spectra before and after TPT point, 
corresponding to their calculated projected bands plotted in Figs.\ref{fig:thetaTPT}e,f.  
At some typical points in the spectra, real-space distributions of electromagnetic fields are imaged by infrared  camera, see examples in Figs.\ref{fig:thetaTPT}g-j. 
Figures \ref{fig:thetaTPT}g and \ref{fig:thetaTPT}h record light distributions before and after TPTs, while Fig.\ref{fig:thetaTPT}i displays an always existing edge mode at bandgaps \ding{183}. In bulk modes, light dissipates into the bulk,  in Fig.\ref{fig:thetaTPT}j.  
Topological immunity against structure defects is tested in Fig.\ref{fig:thetaTPT}k, where 
one cell is removed by adjusting its coupling to the bar state forming a lattice defect. It indicates an unique  ability to withstand and tolerate  structure defects. 
Our topological chip could provide  a fertile ground for studying the critical conditions for the emergence of defect-induced states~\supercite{lustig2022photonic,liu2021bulk,peterson2021trapped}.

\vspace{1.5mm}

$\varphi_{S}$-driven topological phase transitions: 
For typical Floquet TIs, introducing local phase modulations is challenging,  
owning to the globally consistent Floquet period in time domain. 
On our  chip,  Floquet TPTs also can be realised by finely altering the resonant phases ($\varphi_{s}$) in all the site rings. %  is illustrated in Fig.~\ref{fig:TPTphi}. 
In Fig.\ref{fig:TPTphi}a, by turning  $\varphi_{s}$ from 0 to $\pi$, we continue to reduce the number of non-trivial bandgaps in one FSR from two to one in the $\varphi_{s}$-TPT (when we set $\theta= 0.4\pi$ and phase in link rings $\varphi_{L1-L4}=0$). 
Band deformations and changes of topological invariants are shown in the calculated band-structures. 
%, when increasing $\varphi_{s}$ from 0 to $\pi$. 
Bandgaps \ding{184} become trivial after across the critical point $\varphi_{s}=0.58\pi$. As there is a 2$\pi$ period on resonant phases, it is expected that the spectra will return to its original state when $\varphi_{s}=2\pi$. Consequently, there must be another TPT to regenerate topological edge modes at the forbidden bandgaps when increasing $\varphi_{s}$ from $\pi$ to $2\pi$. 
Figure \ref{fig:TPTphi}b displays consistent spectra both in theory and experiment, showing the disappearance of edge states at bandgaps \ding{184} within [0, $\pi$] evolution and the re-emergence  of edge states at bandgaps \ding{182} within [$\pi$, $2\pi$] evolution. 
Interestingly, 
the seemingly negligible $2\pi$ phase in site rings in fact  leads to a reversing of band-structure   and a global phase shift to lower quasi-energy that corresponds to longer wavelength. That being said, the non-trivial bandgaps \ding{183} and \ding{184} at $\varphi_{s}=0$ correspond to non-trivial bandgaps \ding{182} and \ding{183} at $\varphi_{s}=2\pi$, respectively.

Robustness, as the most intriguing property of topological edge modes, allows the protection of  transport immune to imperfections. As long as the presence of disorders does not interrupt the band-structure and the bandgaps keep open, the topological invariants always remain constant and light transport along the edge modes are robust. 
This property has led to many potential applications~\supercite{Hafezi2011,Bandreseaar4005,zeng2020electrically,Mittal2018,BlancoRedondo568,Mittal2021,dai2022topologically,yang2020terahertz,kumar2022phototunable}. 
Previously,  single  or several samples are fabricated, or together with numerical  simulation, to verify   topological robustness. 
By harnessing the individual programmability, we experimentally certificate the robustness of topological edge modes by statistic  measurements. 
 Random perturbations on resonant phases with  uniform distribution of $\delta\times[-0.5,0.5]$ are added to all microrings.  We consider a nontrivial device with an initial configuration of $\varphi_{s} =0.8\pi$, see its band-structure in Fig.\ref{fig:TPTphi}c. 
A set of  100 samples with precisely controlled disorders at $\delta=0.1\pi$ are generated  and tested on a single chip.  The collections of these statistic measurements (gray lines) are shown in Fig.\ref{fig:TPTphi}d. And the spectra for an ideal device with no disorder (blue line) is plot for comparison, in which 
 the topological edge modes in bandgaps \ding{183} are wide and flat high-transmission plateaus. 
In the presence of disorders, the high-transmittance plateaus in topological edge modes exhibit only small fluctuations, but high fluctuations in bulk modes.  
We then estimate the normalised standard deviation of transmittance over 100 samples for different level of disorders, see Fig.\ref{fig:TPTphi}.  
With these statistic measurements, the observation of low-noise windows for topological edge modes unambiguously certificates  the topological robustness against a certain degree of disorders.

Despite the superiority of topological transport, strong disorders may lead to drastic deformation of bands and even disrupt the band topology. But it does not mean the properties of the original topological insulator will completely disappear. 
Interestingly, under specific conditions, the unidirectional transport of the boundary states will still occur in the presence of strong disorders or even amorphous structures~\supercite{stutzer2018photonic,li2009topological,mitchell2018amorphous,jia2023disordered}. Exploring order within disorders is the charm of topology, which exactly requires a highly  programmable platform with individual controllability. 
Recently, the emergence of counterintuitive topological Anderson insulators (TAIs) from trivial phases has been successfully  observed, 
by inducing sufficiently  strong disorders in one sample\supercite{stutzer2018photonic}. 
Similar to Anderson localisation\supercite{Segev2013}, topological Anderson insulating is also a statistical phenomenon for waves in disordered lattices. 
%As a counterintuitive phase, the study of TAIs has been deepening in recent years, 
Such statistical measurement and verification of TAIs  have not been reported in optical systems, to the best of our knowledge. 
 %and statistical results are urgently needed from experiments to prevent the influence of exceptions. 
 %
 Figure \ref{fig:TAI}b illustrates  the random phase distribution in the TAI lattice in the presence of  strong disorders. 
We firstly consider an ideal lattice in the absence of disorders, the coupling $\theta$ is set as 0.3$\pi$, constructing one trivial bandgap within one  FSR. We here interest  in the bandgaps used to be  forbidden, i.e, the dips of blue spectra in Figs.\ref{fig:TAI}c,d. 
Experiment and simulation results of averaged transmission spectra over 100  samples with different levels of disorders in resonant phase are reported  in  Figs.\ref{fig:TAI}c,d, respectively. 
A peak gradually emerges at the windows of forbidden bandgaps (indicated by the red arrow), as the strength of disorders reaches a sufficiently large value, indicating the occurrence of topological Anderson phase transitions.  
The emergence of TAI phase can be portrayed by real-space topological invariants~\supercite{mondragon2014topological,kitaev2006anyons,PhysRevX.3.031005}. 
Analogous to the winding number $\mathcal{W}_{\epsilon}$ in momentum space, the real-space $\mathcal{W}_{real}$ related to non-trivial bandgaps approaches one, while it fluctuates around zero for trivial bandgaps. 
According to the averaged $\mathcal{W}_{real} $ in Fig.~\ref{fig:Wreal_TAI}, a non-zero plateau obviously raises from the ordinary zero dip in forbidden bandgaps. 
Moreover, Fig.\ref{fig:TAI}e shows the imaged real-space field distributions as an increase of disorders, each of which are overlaid distributions of  all 100 samples %in the windows of TAI phase are overlaid 
for better characterising the dynamics of  phase transitions.   
The TAI boundary modes break free from the localisation near the input, and unidirectionally  transport along boundaries with an exponential decay into the bulk lattice. 
In contrast, the same measurements were conducted in a trivial CROW (see inset in Figs.\ref{fig:TAI}f,g).  The shape of spectra remains unchanged and no TAI boundary modes are observed, as shown in Figs.\ref{fig:TAI}f,g.

We further benchmark photonic TIs in various lattice structures.
%On a reconfigurable photonic TI, breaking free from the constraints of lattice structures becomes a relatively simple endeavor. 
Experimental results for the well-known  Su-Schrieffer-Heeger (SSH) 1D TIs are shown in Fig.~\ref{fig:SSH} in Supplementary Information. 
The redundant dimension for the 1D models in a 2D lattice allows observations of the non-Hermitian skin effect in  
Fig.~\ref{fig:NHSE}, 
%Fig.\ref{fig:NHSE}
and other unordinary experiments such as non-reciprocity and next-nearest-neighbor coupling are implementable. 
Moreover, it is also possible to achieve other 2D lattice geometries beyond the inherent square lattice by reprogramming  microrings. 
%selectively closing inter-ring coupling and disrupting ring resonances. % thus achieving a breakthrough in the mindset. 
Figure \ref{fig:honeycomb} illustrates an example of equivalent Floquet TIs in the honeycomb lattice. A perfect correspondence between the measured transmission spectra and simulated projected band-structures  in the strong coupling regime ($\theta = 0.08\pi$) and weak coupling regime ($\theta = 0.24\pi$) are shown  in Figs.\ref{fig:honeycomb}e-h, respectively. When the coupling parameter $\theta$ is larger than $0.19\pi$, TPTs happen at the bandgaps across $\epsilon = \pi/T$ and the flat high-transmission plateau turns into a blocked dip. Distinct real-space field distributions for different modes are shown in Figs.\ref{fig:honeycomb}b-d, including topological edge modes conducting along the honeycomb boundaries, dissipatively distributed bulk modes, and inhibitively forbidden bandgaps.   
By distinguishing the winding number, we observe phase transitions in a five-bulk-band structure. Such multiple non-trivial topological phases in standard honeycomb lattices have  been achieved in another recent work~\supercite{pyrialakos2022bimorphic}, using chain-driven laser-written waveguides. That effectively validates the correctness and reliability of our programmable topological chip.

This work has shown   a flexibly  and rapidly programmable topological photonic chip. Mutlifunctionalities are benchmarked by reprograming the generic chip, including dynamic topological phase transitions, realisations of diverse topological lattices, and implementations of statistical measurement of topological processes. Our generic chip could  be directly  used to discover  topological phases of light and understand exotic phenomena.   
Different to conventional linear-optical circuits with only forward operations of classical~\supercite{Shen2017,Wetzstein2020} and quantum\supercite{Wang2018,bao2023very,} states of light , our device possesses  unique  backwards operations with a lattice of optical resonators and it may provide an alternative solution for  classical~\supercite{pai2023experimentally,mohammadi2019inverse,perez2017multipurpose,bogaerts2020programmable} and quantum~\supercite{Wangreview,Pelucchi2022} information processing and computing tasks. With the ultimate scalability of silicon photonics manufacturing and photonics-electronics co-integration\supercite{Atabaki2018}, topological phases of light can be freely programmed through electronic input, and such programmable topological photonic chips could provide a universal platform for fundamental science and topological technologies.

\printbibliography

\noindent 
\subsection*{Acknowledgements}
{We acknowledge support from the National Natural Science Foundation of China (61975001, 61590933, 11734001, 91950204, 92150302, 11527901),  the Innovation Program for Quantum Science and Technology (2021ZD0301500), the National Key R$\&$D Program of China (2019YFA0308702), Beijing Natural Science Foundation (Z190005, Z220008), and Key R$\&$D Program of Guangdong Province (2018B030329001).}

\subsection*{Authors contributions}
{T.D. and J.W. conceived the project. T.D., A.M and J.M equally contributed to this work. T.D., Y.A, X.H. and J.W. designed the devices. T.D., A.M., X.J., Y.Z. and C.Z. implemented the experiment. T.D., A.M. provided the simulations and performed the theoretical analysis. T.D., A.M., Y.A., B.Z. and J.W. discussed and improved the theoretical results. T.D. designed and established the experiment setup. J.M., Y.Y., Z.L., B.T. and J.L. fabricated the device. Q.G. and J.W. managed the project. T.D., A.M. and J.W. wrote the manuscript. All the authors discussed the results and contributed to the manuscript.}

\subsection*{Competing Interests} The authors declare no competing interests. 

\subsection*{Data availability} 
The data that support the findings of this study are available from the corresponding authors upon reasonable request.

\end{document}